\documentclass[aps,prd,reprint,amsmath,amssymb,nofootinbib,floatfix,longbibliography]{revtex4-2}
\usepackage[T1]{fontenc}
\usepackage{graphicx,bm,booktabs}
\usepackage[colorlinks=true,linkcolor=blue,citecolor=blue,urlcolor=blue]{hyperref}
\usepackage{orcidlink}
\graphicspath{{figures/}}

\newcommand{\dd}{\mathrm{d}}

\newcommand{\NCUa}{Department of physics, Nanchang University, Nanchang, 330031, China}
\newcommand{\NCUb}{Center for Relativistic Astrophysics and High Energy Physics, Nanchang University, Nanchang, 330031, China}

\newcommand{\Gdot}{\textcolor[HTML]{B65F00}{$\bullet$}\,}
\newcommand{\Edot}{\textcolor[HTML]{2C5EAD}{$\bullet$}\,}

\newcommand{\EPlabelj}[6]{%
\mathrm{EP}^{#1,(#6)}_{#2,#3;\{#4,#5\}}%
  }

\usepackage{xcolor}
\usepackage{ulem}

\newif\ifKNchanges
\KNchangestrue
\ifdefined\KNclean
  \KNchangesfalse
\fi

\DeclareRobustCommand{\jkx}[1]{\ifKNchanges{\color[rgb]{0.8,0.4,0}\sout{#1}}\fi}

\DeclareRobustCommand{\EPlabel}[5]{\mathrm{EP}^{\mathrm{#1}}_{#2,#3;\{#4,#5\}}}
\begin{document}
\title{Excitation factors and exceptional points of Kerr--Newman quasinormal modes}

\author{Jun-Kun Zhao \orcidlink{0000-0002-5835-3920}}
%\email{junkunzhao@itp.ac.cn}
\affiliation{Institute of Theoretical Physics, Chinese Academy of Sciences, Beijing 100190, China}

\author{Tieguang Zi\,\orcidlink{0000-0003-0046-2056}}
\email{zitieguang@ncu.edu.cn}
\affiliation{\NCUa}
\affiliation{\NCUb}

\begin{abstract}
We study excitation factors and investigate exceptional points (EPs) of Kerr--Newman quasinormal modes using the exact separated gravitoelectromagnetic equations. At fixed charge, the magnitudes of the excitation factors for the gravitational-led modes studied here decrease with increasing spin, whereas those for electromagnetic-led modes exhibit both monotonic and nonmonotonic behavior. The magnitude of the excitation factor for the first overtone can be comparable to or larger than that for the fundamental mode. We identify 42 EPs at which pairs of modes coalesce in the gravitational-led and electromagnetic-led sectors. Some EPs form sequences involving successive overtone pairs, with higher-overtone pairs coalescing progressively closer to the extremal limit. The additional EPs identified outside these sequences all share the same mode-pair labels as sequence members, but occur at different values of the black hole spin and charge. These findings clarify how rotation and charge shape the relative excitation strengths of different modes and the pattern of mode coalescence in the non-Hermitian spectrum of Kerr--Newman black holes.
\end{abstract}

\maketitle

%%%%%%%%%%%%%%%%%%%%%%%%
\section{Introduction}
%%%%%%%%%%%%%%%%%%%%%%%%
According to the black hole no-hair theorem, stationary, asymptotically flat black holes in four-dimensional Einstein--Maxwell theory belong to the Kerr--Newman (KN) family, characterized by their mass $M$, angular momentum $J=Ma$, and electric charge $Q$~\cite{Israel:1967wq,Carter:1971zc,Robinson:1975bv,Mazur:1982db,Chrusciel:2012jk}. As an exact solution of the Einstein--Maxwell equations, the KN spacetime generalizes both the Kerr and Reissner--Nordstr\"om black holes, providing a framework for investigating the effects of rotation and electric charge on black hole geometry, thermodynamics, and dynamics. From an observational perspective, gravitational-wave observations and black hole imaging offer opportunities to test strong-field gravity and constrain possible deviations from the Kerr geometry~\cite{LIGOScientific:2021sio,EventHorizonTelescope:2022xqj}. Although astrophysical black holes are expected to carry negligible net electric charge, the KN solution remains an important theoretical reference for studying the influence of electric charge on stationary spacetimes. Linear perturbation theory describes the dynamical response of black holes to small disturbances and offers a powerful probe of their geometries~\cite{Berti:2009kk,Konoplya:2011qq}. Their resonant oscillations are characterized by quasinormal modes (QNMs), whose complex frequencies determine the oscillation rates and damping times of black hole ringdown. For given mode labels, these frequencies depend on the black hole background, while the amplitudes of the excited modes also depend on external sources or initial conditions~\cite{Regge:1957td,Vishveshwara:1970zz,Leaver:1985ax,Kokkotas:1999bd,Berti:2009kk}.

Within the Newman--Penrose formalism~\cite{Newman:1961qr}, gravitational and electromagnetic perturbations of Kerr black holes satisfy decoupled, separable Teukolsky equations~\cite{Teukolsky:1973ha,Chandrasekhar:1983bbh}. For KN black holes, however, the background electromagnetic field couples gravitational and electromagnetic perturbations, making their separation considerably more challenging. Previous studies employed the Dudley--Finley approximation~\cite{Dudley:1977zz,Berti:2005eb}, slow-rotation and small-charge expansions~\cite{PaniPRL,Pani:2013wsa,Mark:2014aja}, analytical approximations and WKB methods~\cite{Kokkotas:1993ef,Hod:2008se,Hod:2014uqa,Hod:2015xlh}, and direct numerical solutions of the coupled equations~\cite{Dias:2015wqa,Dias:2021yju,Dias:2022oqm}. These numerical studies of the coupled perturbation equations have provided evidence for linear mode stability and revealed eigenvalue repulsion near extremality.

Recently, Hintz made an important breakthrough by establishing the exact separation of the radiative gravitoelectromagnetic perturbation equations on KN backgrounds~\cite{Hintz:2026jiw}. This resolves a longstanding difficulty in KN perturbation theory by separating the coupled gravitational and electromagnetic perturbations into radial and angular equations linked by a common separation constant $\lambda$. Using these equations, Destounis, Cardoso, and Hintz computed KN QNM frequencies in excellent agreement with previous independent numerical results~\cite{Destounis:2026ljd}. This framework has also been applied to the tidal response of KN black holes~\cite{Chakraborty:2026kyf} and extended to Kerr--Newman--(anti-)de Sitter spacetimes~\cite{Zhao:2026gen,Yoshida:2026yly}.

While QNM frequencies determine the oscillation rates and damping times, they do not characterize the strengths of the corresponding perturbations. Excitation factors (EFs), which are related to the residues of QNM poles, provide complementary information about the source-independent response of black holes. Another important feature of the QNM spectrum is the existence of exceptional points (EPs), where both eigenvalues and their associated eigenfunctions coalesce in such non-Hermitian eigenvalue problems~\cite{Kato:1995cw,Heiss:2012,Miri:2019,Cao:2025afs,Wu:2025wbp,Wu:2026tfx,Rossi:2026als}. Near such degeneracies, QNM frequencies can become highly sensitive to variations in black hole parameters, and the corresponding EFs may exhibit strong enhancements~\cite{Motohashi:2024fwt}. During the preparation of this manuscript, we became aware of two related studies of KN EPs. One identified EPs and eigenstate hysteresis in the coupled gravitoelectromagnetic QNM spectrum of KN black holes~\cite{Arcos:2026eis}. The other investigated spectral instability near KN EPs through pseudospectra and condition numbers for neutral massless scalar fields and coupled gravitoelectromagnetic perturbations~\cite{Destounis:2026qne}. Studying EFs and EPs together therefore provides further insight into the spectral structure and resonant response of black holes.

In this work, we investigate the EFs and EPs of KN QNMs using the exact separated equations for coupled gravitoelectromagnetic perturbations. We first examine how rotation and electric charge affect the EFs of selected gravitational-led and electromagnetic-led modes, comparing their magnitudes and complex-plane trajectories. We then identify 42 EPs in the gravitational-led and electromagnetic-led sectors, revealing ordered overtone ladders and additional off-ladder coalescences. These results provide further insight into the excitation properties and non-Hermitian spectral structure of coupled gravitoelectromagnetic perturbations of KN black holes.

The paper is organized as follows. Section~\ref{sec:equations} introduces the separated perturbation equations, the corresponding boundary conditions, and the methods for computing QNMs, EFs, and EPs. Section~\ref{sec:results} presents the numerical results, including the dependence of EFs on black hole spin and charge and the behavior of QNMs near EPs. Finally, Section~\ref{sec:conclusion} summarizes our main findings and discusses possible future directions.

%%%%%%%%%%%%%%%%%%%%%%%%
\section{Method}\label{sec:equations}
%%%%%%%%%%%%%%%%%%%%%%%%
\subsection{Gravitoelectromagnetic perturbations of Kerr--Newman black holes}
\label{subsec:KN_separated_perturbations}
We consider an asymptotically flat Kerr--Newman black hole in Einstein--Maxwell theory. In geometrized units $G=c=1$ and with metric signature $(+,-,-,-)$, the background metric in Boyer--Lindquist coordinates reads
\begin{equation}
\begin{split}  \label{eq:KN_metric}
\dd s^2= &\frac{\Delta}{\Sigma} \left(\dd t-a\sin^2\theta \dd\phi\right)^2 -\frac{\Sigma}{\Delta}\dd r^2 -\Sigma\dd\theta^2 \\
&-\frac{\sin^2\theta}{\Sigma} \left[a\,\dd t-\left(r^2+a^2\right) \dd\phi\right]^2 \,,
\end{split}
\end{equation}
where
\begin{equation} \label{eq:KN_DeltaSigma}
\Delta=r^2-2Mr+a^2+Q^2 \,,\qquad \Sigma=r^2+a^2\cos^2\theta \,.
\end{equation}
Here $M$, $a=J/M$, and $Q>0$ denote the black hole mass, spin parameter, and electric charge, respectively. We restrict to subextremal configurations, $a^2+Q^2<M^2$, with the outer and inner horizons located at
\begin{equation}
r_{\pm}=M\pm\sqrt{M^2-a^2-Q^2}.
\end{equation}
For convenience, we use the dimensionless parameters $\hat a=a/M$ and $\hat Q=Q/M$. Hintz obtained an exact separation of the radiative gravitoelectromagnetic perturbation equations on the KN background introduced above~\cite{Hintz:2026jiw}. QNM frequencies computed from these equations were found to agree with existing numerical results for the coupled system~\cite{Destounis:2026ljd}. We adopt the separated variables and conventions of these works below.

%%%%%%%%%%%%%%%%%%%%%%%%
\noindent {\bf \emph{Angular eigenvalue problem.}} 
%%%%%%%%%%%%%%%%%%%%%%%%
We take the time and azimuthal dependence of the perturbations to be $e^{-i\omega t+im\phi}$, with $m\in\mathbb Z$. For each spin label $\varsigma=\pm1$, the angular amplitudes $\chi_1(\theta)$ and $\chi_2(\theta)$ satisfy
\begin{equation}
\begin{pmatrix}
T(\theta)&i[\partial_\theta+W(\theta)] \\
i[\partial_\theta-W(\theta)]&-T(\theta)
\end{pmatrix}
\begin{pmatrix}\chi_1\\ \chi_2
\end{pmatrix}
=\lambda \begin{pmatrix}\chi_1\\ \chi_2\end{pmatrix},
\label{eq:KN_angular_matrix}
\end{equation}
where $\lambda$ is the separation constant and $\varsigma=\pm1$ distinguishes the positive- and negative-spin descriptions. The functions $W(\theta)$ and $T(\theta)$ are given by
\begin{equation}
\begin{split}
W(\theta) &=\frac32\cot\theta+\varsigma \left[\frac{m}{\sin\theta}
-\frac{a\omega(M^2+Q^2)}{M^2-Q^2}\sin\theta \right] \,,\\
T(\theta) &=-\frac{3M}{2Q} -\varsigma\frac{2a\omega MQ}{M^2-Q^2}\cos\theta \,.
\end{split}
\label{eq:KN_WT}
\end{equation}

We allow both $\omega$ and $\lambda$ to be complex. Regularity at the poles $\theta=0$ and $\theta=\pi$ requires the angular functions to satisfy
\begin{equation}
\begin{split}
\chi_s(\theta\to0)
&=\theta^{\alpha_s+1/2}
 \left[c_{s,0}+O(\theta^2)\right] \,,\\
\chi_s(\theta\to\pi)
&=(\pi-\theta)^{\beta_s+1/2}
 \left[d_{s,0}+O((\pi-\theta)^2)\right] \,,
\end{split}
\end{equation}
where
\begin{equation}
\alpha_s=|\varsigma m+s|,\qquad \beta_s=|\varsigma m-s|,\qquad s=1,2 \,.
\label{eq:KN_pole_exponents}
\end{equation}
Here $s=1,2$ labels the two angular components. For a given frequency $\omega$, the angular equation subject to these boundary conditions determines the allowed eigenvalues $\lambda$.

At $a=0$, the angular eigenvalues are known explicitly. For $\ell\ge2$ and $|m|\le\ell$, they are
\begin{equation}
\lambda^{(0)}_{\ell b} =b\sqrt{(\ell-1)(\ell+2)+\frac{9M^2}{4Q^2}} \,, \qquad b=\pm1 \,.
\label{eq:KN_lambda_seed}
\end{equation}
For $\ell=1$, when only the $b=-1$ angular eigenvalue can be allowed, $\lambda^{(0)}_{1,-}=-3M/(2Q)$, with $m=-1,0,1$. As $a$ is varied from zero, we follow each angular eigenvalue with labels $(\ell,b)$.
The modes with $b=+1$ and $b=-1$ reduce to the gravitational and electromagnetic perturbations of Kerr black hole , respectively, as black hole charge $Q\to0$. We therefore refer to them as gravitational-led and electromagnetic-led modes. The equations written above contain terms proportional to $1/Q$, so this limit cannot be obtained by directly setting $Q=0$. The required variable rescalings are given in~\cite{Destounis:2026ljd} and summarized in Appendix~\ref{app:Q-to-0}.

%%%%%%%%%%%%%%%%%%%%%%%%
\noindent {\bf \emph{Radial equation.}} 
%%%%%%%%%%%%%%%%%%%%%%%%
To write down the radial equations, we first define
\begin{align}
K(r)&=-(r^2+a^2)\omega+am \,,\quad \xi_0=\frac{2\omega Q^2}{M^2-Q^2},\nonumber\\
\widetilde K(r)&=K(r)-\xi_0\Delta(r) \,,\\
R_1^{(\varsigma)}(r) &=-\varsigma\frac{2\omega Q(Mr-Q^2)}{M^2-Q^2}
+\frac{3iM}{2Q},\nonumber\\
h(r)&=R_1^{(\varsigma)}(r)+i\varsigma\lambda.
\label{eq:KN_R1h}
\end{align}
Following Ref.~\cite{Destounis:2026ljd}, we introduce the radial functions
\begin{equation}
\begin{aligned}
w^{(+)}&=\Delta^{3/4}H_r^{-1}e^{i\xi_0r}z_2^{(+)},\\
w^{(-)}&=\Delta^{-3/4}H_r^{-1}e^{i\xi_0r}z_1^{(-)},
\end{aligned}
\label{eq:KN_radial_transform}
\end{equation}
where $z_{1,2}^{(\varsigma)}$ are the radial amplitudes in the separated first-order equations and
\begin{equation}
H_r(r)=\exp\left[i\int_{2r_+}^{r}
\frac{K(\widetilde r)}{\Delta(\widetilde r)}
\,d\widetilde r\right].
\label{eq:KN_Hr}
\end{equation}

For each $\varsigma=\pm1$, the radial function $w=w^{(\varsigma)}$ satisfies the following second-order equation wherever $h(r)\neq0$
\begin{equation}
\begin{split}
\Biggl\{& \left[ \partial_r -\frac{(1+3\varsigma)\Delta'}{2\Delta}
+\frac{2i\widetilde K}{\Delta} -\frac{(R_1^{(\varsigma)})'}{h} \right]\Delta\partial_r \\
&-\left[(R_1^{(\varsigma)})^2+\lambda^2\right] \Biggr\}w=0 \,,
\end{split}
\label{eq:KN_radial_factored}
\end{equation}
where primes denote derivatives with respect to $r$. Expanding the derivatives and multiplying by $h(r)$, we obtain the equivalent equation
\begin{equation} \label{eq:KN_radial_ABC}
A(r)w''+B(r)w'+C(r)w=0,
\end{equation}
with
\begin{equation}
\begin{split}
A&=\Delta h \,, \quad B=\frac{1-3\varsigma}{2}\Delta'h
+2i\widetilde K h-\Delta(R_1^{(\varsigma)})' \,, \\
C&=-h\left[(R_1^{(\varsigma)})^2+\lambda^2\right] \,.
\end{split}
\label{eq:KN_radial_coefficients}
\end{equation}
This form contains no explicit $1/h$ terms and is convenient for numerical calculations.

%%%%%%%%%%%%%%%%%%%%%%%%
\subsection{Boundary conditions and quasinormal modes}\label{subsec:sol}
%%%%%%%%%%%%%%%%%%%%%%%%
We now impose boundary conditions on the separated perturbation equations. The radial solutions are required to be purely ingoing at the event horizon and purely outgoing at spatial infinity. These conditions determine the complex QNM frequencies. We denote two solutions of the radial equation~\eqref{eq:KN_radial_ABC} for $w$ by $R_{\ell m}^{\mathrm{in}}(r)$ and $R_{\ell m}^{\mathrm{up}}(r)$, with the usual in/up notation~\cite{Motohashi:2021zyv}. The first is characterized by purely ingoing behavior at the event horizon, while the second is characterized by purely outgoing behavior at infinity. For generic frequencies, the two solutions are linearly independent and have the following asymptotic forms
\begin{align}
R^{\mathrm{in}}_{\ell m}(r)
&\simeq
\begin{cases}
A_{\ell m}^{(\mathrm{trans})}g_H^{\mathrm{in}}(r) \,, & r\to r_+ \,,\\
A_{\ell m}^{(\mathrm{ref})}f_\infty^{\mathrm{out}}(r)+A_{\ell m}^{(\mathrm{inc})}f_\infty^{\mathrm{in}}(r) \,, & r\to\infty \,,
\end{cases}
\label{eq:Rin}\\
R^{\mathrm{up}}_{\ell m}(r)
&\simeq
\begin{cases}
C_{\ell m}^{(\mathrm{up})}g_H^{\mathrm{out}}(r) +C_{\ell m}^{(\mathrm{ref})}g_H^{\mathrm{in}}(r) \,, & r\to r_+,\\
C_{\ell m}^{(\mathrm{trans})}f_\infty^{\mathrm{out}}(r)\,, & r\to\infty \,.
\end{cases}
\end{align}
The functions $g_H^{\mathrm{in/out}}$ and $f_\infty^{\mathrm{in/out}}$ describe the ingoing and outgoing solutions near the horizon and at infinity, respectively. The coefficients $A_{\ell m}^{(\cdot)}$ and $C_{\ell m}^{(\cdot)}$ depend on $\omega$ and $\lambda$.

Near the event horizon, we introduce $x=1-r_+/r$. For $h(r_+)\ne0$ and $\rho_H\notin\mathbb Z$, the two independent solutions have the following forms
\begin{equation}
\begin{split}
g_H^{\mathrm{in}}(r) &=1+O(x) \,,\\
g_H^{\mathrm{out}}(r) &=x^{\rho_H}\left[1+O(x)\right] \,,
\end{split}
\end{equation}
where
\begin{equation}
\rho_H=\frac{1+3\varsigma}{2} +\frac{i(\omega-m\Omega_H)}{\kappa_H} \,.
\end{equation}
Here $\Omega_H=a/(r_+^2+a^2)$ and $\kappa_H=(r_+-r_-)/[2(r_+^2+a^2)]$ are the angular velocity and surface gravity of the event horizon, respectively. The transformation defining $w$ removes the ingoing phase and the associated powers of $\Delta$, so the ingoing solution corresponds to the regular branch $g_H^{\mathrm{in}}$. At spatial infinity, for $\omega\ne0$, the ingoing and outgoing solutions behave as
\begin{equation}
\begin{split}
f_\infty^{\mathrm{in}}(r) &=e^{i\xi_0r} (\frac{r}{M})^{3\varsigma} \left[1+\frac{a_1^{\mathrm{in}}}{r} +O(\frac{1}{r^2})\right] \,,\\
f_\infty^{\mathrm{out}}(r) &=e^{i(2\omega+\xi_0)r}(\frac{r}{M})^{4iM\omega} \left[1+\frac{a_1^{\mathrm{out}}}{r} +O(\frac{1}{r^2})\right] \,.
\end{split}
\end{equation}
The coefficients $a_1^{\mathrm{in/out}}$ and the higher-order terms are determined by substituting these expansions into the radial equation.

The QNM frequencies $\omega_{\ell mn}$ are obtained by requiring the incoming wave amplitude of $R_{\ell m}^{\mathrm{in}}(r)$ to vanish at infinity
\begin{equation}
A_{\ell m}^{(\mathrm{inc})}
\bigl(\omega_{\ell mn}, \lambda(\omega_{\ell mn})\bigr)=0 \,.
\end{equation}
Here $\lambda(\omega)$ is the angular eigenvalue satisfying the regularity conditions on the chosen $(\ell,m,b)$ branch, and $n$ denotes the radial overtone number. The labels $b$ and $\varsigma$ are suppressed for simplicity.

The QNMs $\omega_{\ell mn}$ and separation constant $\lambda$ are determined simultaneously using angular Galerkin projection and radial Chebyshev collocation (see,~\emph{e.g.}~\cite{Destounis:2026ljd}). Similar techniques have been employed in studies of Kerr angular eigenvalue problems~\cite{Cook:2014cta,Jansen:2017oag}, and coupled gravitoelectromagnetic perturbations of KN black holes~\cite{Dias:2015wqa,Dias:2022oqm,Pombo:2026xsj}. The numerical procedure is described in Appendices~\ref{app:angular-basis}--\ref{app:joint-eigenvalues}.

%%%%%%%%%%%%%%%%%%%%%%%%
\subsection{Excitation factors and exceptional points in QNM spectrum}\label{sec:exceptional-points}
%%%%%%%%%%%%%%%%%%%%%%%%

In addition to the QNM frequencies, the EFs characterize the source-independent strength of the corresponding perturbations. For the separated radial variable $w$, they are determined by the outgoing amplitude $A_{\rm ref}$ and the frequency derivative of the incoming amplitude $A_{\rm inc}$, with the boundary conditions specified in Sec.~\ref{subsec:sol}.

For a simple QNM with $dA_{\rm inc}/d\omega|_{\omega_q}\ne0$, the excitation factor is defined in terms of the asymptotic amplitudes in~\eqref{eq:Rin}:
\begin{equation}\label{eq:excitation_factor}
\mathcal B_{lmn}^{(w)} =\frac{A_{\ell m }^{(\text{ref})}}{2 \omega_{\ell m n}}\left(\frac{dA^{(\rm inc)}_{\ell m}}{d\omega}\right)^{-1}_{\omega=\omega_{\ell mn}}\,.
\end{equation}
The outgoing amplitude in the numerator is evaluated at $\omega=\omega_{\ell mn}$, and the frequency derivative includes the dependence of the angular eigenvalue $\lambda$ on $\omega$, with $a$ and $Q$ held fixed. The amplitude of a QNM in a physical ringdown signal depends on the external source or initial conditions, and its calculation requires reconstructing the corresponding radiation fields from the separated variables. In contrast, the EF $\mathcal B_{lmn}^{(w)}$ is determined by the homogeneous perturbation equations and their boundary conditions, independently of the source.

The ingoing and outgoing boundary conditions at the event horizon and infinity make the QNM eigenvalue problem non-self-adjoint, allowing for the existence of EPs. At an EP, two QNM frequencies and their associated eigenfunctions coalesce. For Kerr--Newman black holes, the rotation parameter $a$ and electric charge $Q$ provide two independent parameters for exploring such degeneracies.

For the fixed black hole  parameters $(a,Q)$ and a chosen angular eigenvalue branch $\lambda(\omega;a,Q)$, the incoming-wave amplitude is defined as
\begin{equation}
\mathcal{A}(\omega;a,Q)
=A_{\rm inc}\bigl(\omega,\lambda(\omega;a,Q);a,Q\bigr) \,.
\label{eq:incoming-amplitude}
\end{equation}
The QNM frequencies correspond to the zeros of $\mathcal A$. A simple QNM at $\omega=\omega_q$ satisfies $\mathcal A(\omega_q)=0$ and $D_\omega\mathcal A(\omega_q)\ne0$, where $D_\omega$ denotes differentiation along the angular eigenvalue branch. As the black hole parameters vary, two QNM frequencies may coalesce, producing a double zero of $\mathcal A$. A candidate second-order EP therefore satisfies
\begin{equation}
\mathcal{A}=0 \,, \qquad
D_\omega\mathcal{A}=0 \,, \qquad
D_\omega^2\mathcal{A}\neq0 \,,
\label{eq:double-zero}
\end{equation}
evaluated at $(\omega_{\rm EP},a_{\rm EP},Q_{\rm EP})$. The frequency derivative is taken along the angular eigenvalue branch $\lambda(\omega;a,Q)$, with $a$ and $Q$ held fixed:
\begin{equation}
D_\omega\mathcal{A} =\frac{\partial A_{\rm inc}}{\partial\omega} +\frac{\partial A_{\rm inc}}{\partial\lambda} \frac{\partial\lambda}{\partial\omega}   \,.
\label{eq:Doemga}
\end{equation}

The connection between EPs and EFs follows from~\eqref{eq:excitation_factor}, which contains the inverse frequency derivative $(D_\omega\mathcal A)^{-1}$ for a simple QNM. As two QNM frequencies approach an EP, this derivative in Eq.~\eqref{eq:Doemga} tends to zero. Similar discussions of EPs near avoided crossings have been observed in Kerr QNMs~\cite{Motohashi:2024fwt}. The accuracy of the computed QNM and EP frequencies is assessed by comparison with previous results in Appendix~\ref{app:numer:error}.

%%%%%%%%%%%%%%%%%%%%%%%%
\section{Results}\label{sec:results}
\subsection{Spin and charge dependence of the excitation factors}\label{sec:maps}
\begin{figure}[!htbp]
\centering
\includegraphics[width=\columnwidth]{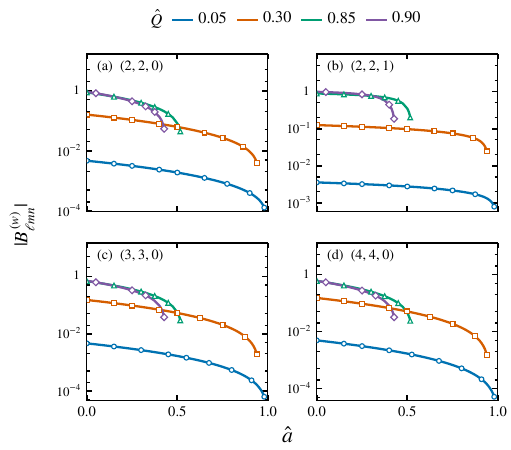}
\caption{Absolute values of the gravitational-led EFs, $|\mathcal B_{\ell mn}^{(w)}|$, as functions of the dimensionless black hole spin $\hat a$. Panels (a)--(d) correspond to $(\ell,m,n)=(2,2,0)$, $(2,2,1)$, $(3,3,0)$, and $(4,4,0)$, respectively. In each panel, the curves correspond to the fixed charges
$\hat Q=0.05$, $0.30$, $0.85$, and $0.90$, and extend from $\hat a=0$ toward their respective extremal bounds.}\label{fig:Gmap}
\end{figure}

\begin{figure}[!htbp]
\centering
\includegraphics[width=\columnwidth]{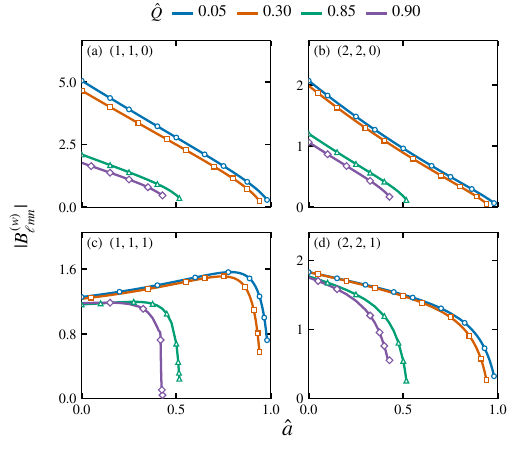}
\caption{Absolute values of the electromagnetic-led EFs, $|\mathcal B_{\ell mn}^{(w)}|$, as functions of the dimensionless black hole spin $\hat a$. Panels (a)--(d) correspond to $(\ell,m,n)=(1,1,0)$, $(2,2,0)$, $(1,1,1)$, and $(2,2,1)$, respectively. In each panel, the curves correspond to the fixed charges $\hat Q=0.05$, $0.30$, $0.85$, and $0.90$, and extend from $\hat a=0$ toward their respective extremal bounds.}\label{fig:EMmap}
\end{figure}

%%%%%%%%%%%%%%%%%%%%%%%%
We investigate how rotation and electric charge affect the EFs of KN QNMs. Figures~\ref{fig:Gmap} and~\ref{fig:EMmap} present the gravitational-led ($b=+1$) and electromagnetic-led ($b=-1$) sectors, respectively. In the gravitational-led sector, we consider $(\ell,m,n)=(2,2,0)$, $(2,2,1)$, $(3,3,0)$, and $(4,4,0)$, whereas the electromagnetic-led sector includes $(1,1,0)$, $(1,1,1)$, $(2,2,0)$, and $(2,2,1)$. The curves are shown for four fixed charges, $\hat Q=0.05$, $0.30$, $0.85$, and $0.90$, with the black hole spin $\hat a$ varied from zero toward the extremal bound $\hat a=\sqrt{1-\hat Q^2}$, while remaining in the subextremal region $\hat a^2+\hat Q^2<1$.

\paragraph{Gravitational-led sector.} As shown in Fig.~\ref{fig:Gmap}, the EF magnitudes of all four selected gravitational-led modes decrease overall as the black hole spin increases at fixed charge. The strength of this suppression, however, depends strongly on the mode. Table~\ref{tab:EF_spin_suppression} quantifies this difference for the fundamental and first-overtone modes in the $\ell=m=2$ sector at $\hat Q=0.05$. As $\hat a$ increases from $0$ to $0.98$, $|\mathcal B_{220}^{(w)}|$ decreases by a factor of approximately $36$, whereas $|\mathcal B_{221}^{(w)}|$ decreases by a factor of only about $4.5$.

The stronger suppression of the fundamental mode reverses the relative ordering of the two EFs: $|\mathcal B_{220}^{(w)}|$ is larger at $\hat a=0$, while $|\mathcal B_{221}^{(w)}|$ is larger at $\hat a=0.98$. Thus, black hole rotation can reorganize the relative excitation strengths of different overtones even within the same angular sector. The $(3,3,0)$ and $(4,4,0)$ modes show the same overall decrease with spin, although their rates of suppression are different. Figure~\ref{fig:Gmap} also reveals a nontrivial charge dependence. Over the spin interval covered by all four curves, the EF magnitudes generally increase as $\hat Q$ rises from $0.05$ to $0.30$. At higher charge, however, the $\hat Q=0.85$ and $0.90$ curves may cross, reversing their relative ordering as the spin varies.
%%%%%%%%%%%%%%%%%%%%%%%%%%%
\begin{table}[!htbp]
\caption{Gravitational-led EFs at fixed charge $\hat Q=0.05$. The suppression factor is defined by $\mathcal S=|\mathcal B_{\ell mn}^{(w)}(\hat a=0)|/|\mathcal B_{\ell mn}^{(w)}(\hat a=0.98)|$.}\label{tab:EF_spin_suppression}
\centering
\begin{ruledtabular}
\begin{tabular}{cccc}
$(\ell,m,n)$ & $\hat a=0$ & $\hat a=0.98$ & $\mathcal S$ \\ \hline
$(2,2,0)$ & $4.69\times10^{-3}$ & $1.29\times10^{-4}$ & $36.4$ \\
$(2,2,1)$ & $3.57\times10^{-3}$ & $7.92\times10^{-4}$ & $4.51$
\end{tabular}
\end{ruledtabular}
\end{table}
%%%%%%%%%%%%%%%%%%%%%%%%%%%

\paragraph{Electromagnetic-led sector.} As shown in Fig.~\ref{fig:EMmap}, the spin dependence of the electromagnetic-led modes shares some features with, but also differs from, that of the gravitational-led sector. At fixed charge, the EF magnitudes of the $(1,1,0)$, $(2,2,0)$, and $(2,2,1)$ modes decrease overall as the black hole spin increases. The $(1,1,1)$ mode, however, exhibits a nonmonotonic spin dependence: its EF magnitude initially increases, reaches a maximum at relatively large spin, and then decreases as the corresponding extremal bound is approached. The charge dependence is also mode dependent. Over the range of black hole spin considered for all four charges, the EF magnitudes of the $(1,1,0)$, $(2,2,0)$, and $(2,2,1)$ modes generally decrease as the charge increases. The $(1,1,1)$ mode again behaves differently: the $\hat Q=0.85$ and $0.90$ curves cross, so their relative ordering reverses as the spin varies.

\begin{figure*}[!htbp]
\centering
\includegraphics[width=0.85\textwidth]{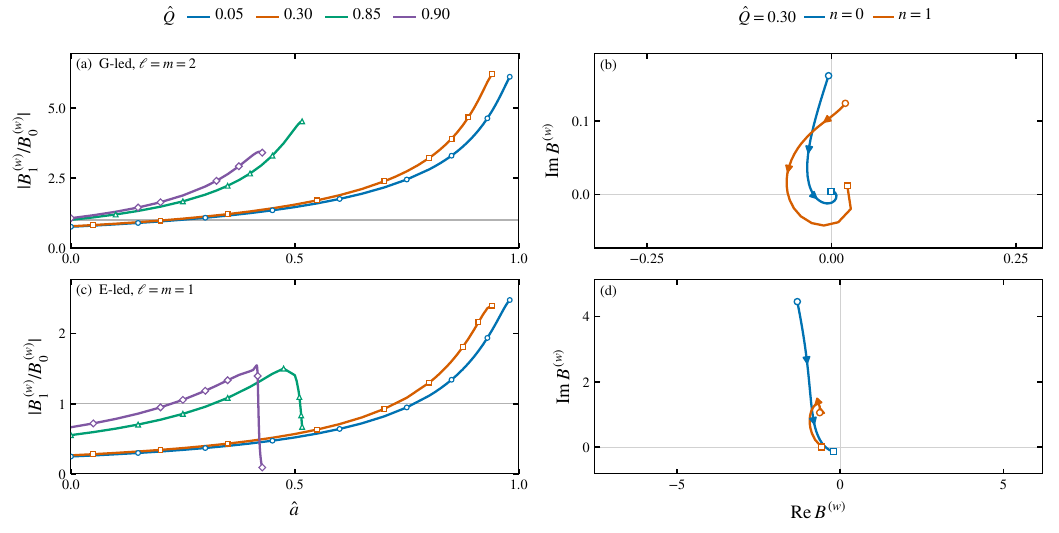}
\caption{Comparison of the fundamental mode ($n=0$) and first overtone ($n=1$) in the gravitational-led $\ell=m=2$ sector (top row) and electromagnetic-led $\ell=m=1$ sector (bottom row). The left column shows $|B^{(w)}_1/B^{(w)}_0|$ versus $\hat a$ for four fixed charges, while the right column shows the complex-plane trajectories at $\hat Q=0.30$; arrows indicate increasing spin.}\label{fig:complexEF}
\end{figure*}

Figure~\ref{fig:complexEF} compares the fundamental mode ($n=0$) and the first overtone ($n=1$) in the gravitational-led $\ell=m=2$ and electromagnetic-led $\ell=m=1$ sectors. The left panels show the ratio $|\mathcal B_1^{(w)}/\mathcal B_0^{(w)}|$ as a function of $\hat a$ at fixed charges. The right panels display the corresponding trajectories of $\mathcal B_0^{(w)}$ and $\mathcal B_1^{(w)}$ in the complex plane at $\hat Q=0.30$, with the arrows indicating increasing spin. For the low-charge cases $\hat Q=0.05$ and $0.30$, the ratio increases with spin in both perturbation sectors and eventually exceeds $1$. Thus, at sufficiently rapid rotation, the EF magnitude of the first overtone becomes larger than that of the fundamental mode. The gravitational-led sector shows a similar increase at the higher charges considered here. In the electromagnetic-led sector, however, the $\hat Q=0.85$ and $0.90$ curves turn over and decrease sharply as their respective extremal bounds are approached, so that the fundamental-mode EF again becomes larger than the first-overtone EF. At $\hat Q=0.30$, the complex-plane trajectories in Figs.~3(b) and~3(d) provide phase information that is absent from the magnitudes alone. In the gravitational-led sector, the fundamental mode and first overtone trace pronounced loops as the black hole spin increases. The electromagnetic-led trajectories exhibit less pronounced winding, but the two modes still follow distinct paths.

%%%%%%%%%%%%%%%%%%%%%%%%
\subsection{Exceptional points of Kerr--Newman gravitoelectromagnetic perturbations}\label{sec:EPs}
%%%%%%%%%%%%%%%%%%%%%%%%
EPs provide a direct window into the non-Hermitian spectral structure of QNM problems. In the Kerr QNM spectrum, varying the single dimensionless spin parameter $a/M$ can lead to avoided crossings and eigenvalue repulsion between QNM branches~\cite{Motohashi:2024fwt,Lo:2025njp}. A generic EP, however, typically requires more than one independent control parameter. The KN family naturally supplies a two-dimensional parameter space through $\hat a$ and $\hat Q$, thereby offering a natural platform for investigating the exceptional-point structure of its QNM spectrum.

Using the conditions introduced in Sec.~\ref{sec:exceptional-points}, we identify 42 EPs in the coupled gravitoelectromagnetic QNM spectrum, as summarized in Table~\ref{tab:EPs}. Twenty-three belong to the gravitational-led sector and nineteen to the electromagnetic-led sector. The gravitational-led EPs occur in the $(\ell,m)=(2,2)$, $(3,2)$, $(3,3)$, and $(4,3)$ modes, whereas the electromagnetic-led EPs are found in the $(1,1)$, $(2,1)$, and $(2,2)$ modes. Additional EPs may exist for other mode pairs or in regions of parameter space not covered by the present search. We denote each EP by
\begin{equation}
\mathrm{EP}^{X,(j)}_{\ell,m; \{n_1,n_2\}} \,,
\end{equation}
where $X=G$ or $E$ denotes the gravitational-led or electromagnetic-led sector, respectively, $(\ell,m)$ gives the angular mode labels, and $(n_1,n_2)$ identifies the two overtone branches whose QNM frequencies and eigenfunctions coalesce. If multiple EPs occur for the same sector and mode pair, the additional index $j$ distinguishes them.

\begin{table*}[!th]
\caption{Parameters and complex frequencies of 42 selected Kerr--Newman QNM EPs, labeled by $\mathrm{EP}^{X,(j)}_{\ell,m;\{n_1,n_2\}}$ and ordered by $(\ell,m,n_1,n_2)$ within each sector. For repeated labels, $j=0$ denotes the selected sequence member, while $j=1,2,\ldots$ orders the other locations by increasing $\hat a_{\rm EP}$; otherwise, $j$ is omitted. Here $d_{\rm EP}=\hat a_{\rm EP}^{\,2}+\hat Q_{\rm EP}^{\,2}-1<0$, with $d_{\rm EP}\to0^{-}$ at extremality.}\label{tab:EPs}
\centering
\small
\setlength{\tabcolsep}{5pt}
\renewcommand{\arraystretch}{1.15}
\begin{ruledtabular}
\begin{tabular}{lrrcr}
EP label
& $\hat a_{\rm EP}$
& $\hat Q_{\rm EP}$
& $M\omega_{\rm EP}$
& $d_{\rm EP}$
\\
\hline

% \multicolumn{5}{l}{\textit{Gravitational-led sector ($b=+1$)}}
% \\

\Gdot $\EPlabel{G}{2}{2}{0}{1}$
& 0.304116510
& 0.952362666
& $0.564398227-0.028304554\,\mathrm{i}$
& $-5.185\times10^{-4}$
\\

\Gdot $\EPlabelj{G}{2}{2}{1}{2}{0}$
& 0.309806527
& 0.950727198
& $0.570215504-0.025135809\,\mathrm{i}$
& $-1.377\times10^{-4}$
\\

\Gdot $\EPlabelj{G}{2}{2}{1}{2}{1}$
& 0.209344101
& 0.971405377
& $0.449330846-0.191881258\,\mathrm{i}$
& $-1.255\times10^{-2}$
\\

\Gdot $\EPlabelj{G}{2}{2}{1}{2}{2}$
& 0.220095829
& 0.972984994
& $0.454477546-0.185196350\,\mathrm{i}$
& $-4.858\times10^{-3}$
\\

\Gdot $\EPlabelj{G}{2}{2}{2}{3}{0}$
& 0.312388626
& 0.949922517
& $0.572850039-0.023703417\,\mathrm{i}$
& $-6.056\times10^{-5}$
\\

\Gdot $\EPlabelj{G}{2}{2}{2}{3}{1}$
& 0.225990537
& 0.972819474
& $0.457397203-0.181809301\,\mathrm{i}$
& $-2.551\times10^{-3}$
\\

\Gdot $\EPlabel{G}{2}{2}{3}{4}$
& 0.313876728
& 0.949446103
& $0.574366343-0.022870044\,\mathrm{i}$
& $-3.350\times10^{-5}$
\\

\Gdot $\EPlabel{G}{2}{2}{4}{5}$
& 0.314850778
& 0.949130064
& $0.575357657-0.022319723\,\mathrm{i}$
& $-2.111\times10^{-5}$
\\

\Gdot $\EPlabelj{G}{2}{2}{5}{6}{0}$
& 0.315541023
& 0.948904312
& $0.576059338-0.021926955\,\mathrm{i}$
& $-1.447\times10^{-5}$
\\

\Gdot $\EPlabelj{G}{2}{2}{5}{6}{1}$
& 0.749057505
& 0.532397483
& $0.555905299-0.791341725\,\mathrm{i}$
& $-1.555\times10^{-1}$
\\

\Gdot $\EPlabel{G}{3}{2}{1}{2}$
& 0.743448001
& 0.667234185
& $0.972691674-0.081879676\,\mathrm{i}$
& $-2.084\times10^{-3}$
\\

\Gdot $\EPlabel{G}{3}{2}{2}{3}$
& 0.761028839
& 0.647930594
& $0.975212331-0.076436630\,\mathrm{i}$
& $-1.021\times10^{-3}$
\\

\Gdot $\EPlabel{G}{3}{2}{3}{4}$
& 0.773694655
& 0.633079754
& $0.977194724-0.073501994\,\mathrm{i}$
& $-6.066\times10^{-4}$
\\

\Gdot $\EPlabel{G}{3}{2}{4}{5}$
& 0.783206434
& 0.621438856
& $0.978681255-0.071596994\,\mathrm{i}$
& $-4.014\times10^{-4}$
\\

\Gdot $\EPlabel{G}{3}{2}{5}{6}$
& 0.790616814
& 0.612078464
& $0.979817398-0.070240484\,\mathrm{i}$
& $-2.850\times10^{-4}$
\\

\Gdot $\EPlabel{G}{3}{3}{0}{1}$
& 0.377451425
& 0.925982662
& $0.994079166-0.012592716\,\mathrm{i}$
& $-8.653\times10^{-5}$
\\

\Gdot $\EPlabel{G}{3}{3}{1}{2}$
& 0.380844649
& 0.924626143
& $0.999684309-0.010783874\,\mathrm{i}$
& $-2.385\times10^{-5}$
\\

\Gdot $\EPlabelj{G}{3}{3}{2}{3}{0}$
& 0.382475757
& 0.923959822
& $1.002389863-0.009969328\,\mathrm{i}$
& $-1.054\times10^{-5}$
\\

\Gdot $\EPlabelj{G}{3}{3}{2}{3}{1}$
& 0.299535626
& 0.952983464
& $0.849371999-0.135671548\,\mathrm{i}$
& $-2.101\times10^{-3}$
\\

\Gdot $\EPlabel{G}{3}{3}{3}{4}$
& 0.383445163
& 0.923560493
& $1.004002063-0.009496059\,\mathrm{i}$
& $-5.822\times10^{-6}$
\\

\Gdot $\EPlabel{G}{4}{3}{1}{2}$
& 0.667095777
& 0.744442401
& $1.394804674-0.058219191\,\mathrm{i}$
& $-7.887\times10^{-4}$
\\

\Gdot $\EPlabelj{G}{4}{3}{2}{3}{0}$
& 0.674811449
& 0.737727892
& $1.398473807-0.053856357\,\mathrm{i}$
& $-3.871\times10^{-4}$
\\

\Gdot $\EPlabelj{G}{4}{3}{2}{3}{1}$
& 0.504647151
& 0.860790519
& $1.244825997-0.196167093\,\mathrm{i}$
& $-4.371\times10^{-3}$
\\

\hline
% \multicolumn{5}{l}{\textit{Electromagnetic-led sector ($b=-1$)}}
% \\

\Edot $\EPlabel{E}{1}{1}{0}{1}$
& 0.687265135
& 0.726012494
& $0.472653256-0.023562859\,\mathrm{i}$
& $-5.725\times10^{-4}$
\\

\Edot $\EPlabel{E}{1}{1}{1}{2}$
& 0.706931140
& 0.707171968
& $0.475069610-0.020331190\,\mathrm{i}$
& $-1.562\times10^{-4}$
\\

\Edot $\EPlabelj{E}{1}{1}{2}{3}{0}$
& 0.716368444
& 0.697672665
& $0.476161256-0.018872304\,\mathrm{i}$
& $-6.911\times10^{-5}$
\\

\Edot $\EPlabelj{E}{1}{1}{2}{3}{1}$
& 0.424335534
& 0.901822747
& $0.394681893-0.198430498\,\mathrm{i}$
& $-6.655\times10^{-3}$
\\

\Edot $\EPlabel{E}{1}{1}{3}{4}$
& 0.721974494
& 0.691892016
& $0.476790542-0.018025020\,\mathrm{i}$
& $-3.827\times10^{-5}$
\\

\Edot $\EPlabel{E}{1}{1}{4}{5}$
& 0.725714231
& 0.687978740
& $0.477202514-0.017466386\,\mathrm{i}$
& $-2.411\times10^{-5}$
\\

\Edot $\EPlabel{E}{2}{1}{2}{3}$
& 0.845724565
& 0.495798164
& $0.558020601-0.321553435\,\mathrm{i}$
& $-3.893\times10^{-2}$
\\

\Edot $\EPlabel{E}{2}{1}{3}{4}$
& 0.893217748
& 0.425148005
& $0.547326088-0.310235376\,\mathrm{i}$
& $-2.141\times10^{-2}$
\\

\Edot $\EPlabelj{E}{2}{1}{4}{5}{0}$
& 0.928040496
& 0.353599541
& $0.540220533-0.303287275\,\mathrm{i}$
& $-1.371\times10^{-2}$
\\

\Edot $\EPlabelj{E}{2}{1}{4}{5}{1}$
& 0.594581447
& 0.783258533
& $0.529534752-0.547071342\,\mathrm{i}$
& $-3.298\times10^{-2}$
\\

\Edot $\EPlabel{E}{2}{1}{5}{6}$
& 0.954235797
& 0.282599155
& $0.534993813-0.298429976\,\mathrm{i}$
& $-9.572\times10^{-3}$
\\

\Edot $\EPlabel{E}{2}{2}{0}{1}$
& 0.653891728
& 0.756531816
& $0.918373632-0.010310056\,\mathrm{i}$
& $-8.522\times10^{-5}$
\\

\Edot $\EPlabelj{E}{2}{2}{1}{2}{0}$
& 0.662380552
& 0.749151818
& $0.922230468-0.008685365\,\mathrm{i}$
& $-2.356\times10^{-5}$
\\

\Edot $\EPlabelj{E}{2}{2}{1}{2}{1}$
& 0.461271757
& 0.883991733
& $0.793226261-0.144859314\,\mathrm{i}$
& $-5.787\times10^{-3}$
\\

\Edot $\EPlabelj{E}{2}{2}{2}{3}{0}$
& 0.666543573
& 0.745459097
& $0.924092978-0.007959284\,\mathrm{i}$
& $-1.040\times10^{-5}$
\\

\Edot $\EPlabelj{E}{2}{2}{2}{3}{1}$
& 0.471332566
& 0.880489277
& $0.795394060-0.137484232\,\mathrm{i}$
& $-2.584\times10^{-3}$
\\

\Edot $\EPlabelj{E}{2}{2}{3}{4}{0}$
& 0.669046534
& 0.743216659
& $0.925204529-0.007539172\,\mathrm{i}$
& $-5.733\times10^{-6}$
\\

\Edot $\EPlabelj{E}{2}{2}{3}{4}{1}$
& 0.477535991
& 0.877784636
& $0.797033638-0.133556710\,\mathrm{i}$
& $-1.454\times10^{-3}$
\\

\Edot $\EPlabel{E}{2}{2}{4}{5}$
& 0.670728499
& 0.741700537
& $0.925948100-0.007262762\,\mathrm{i}$
& $-3.594\times10^{-6}$
\\

\end{tabular}
\end{ruledtabular}
\end{table*}

%%%%%%%%%%%%%%%%%%%%%
\begin{figure*}[!t]
\centering
\includegraphics[width=\textwidth]{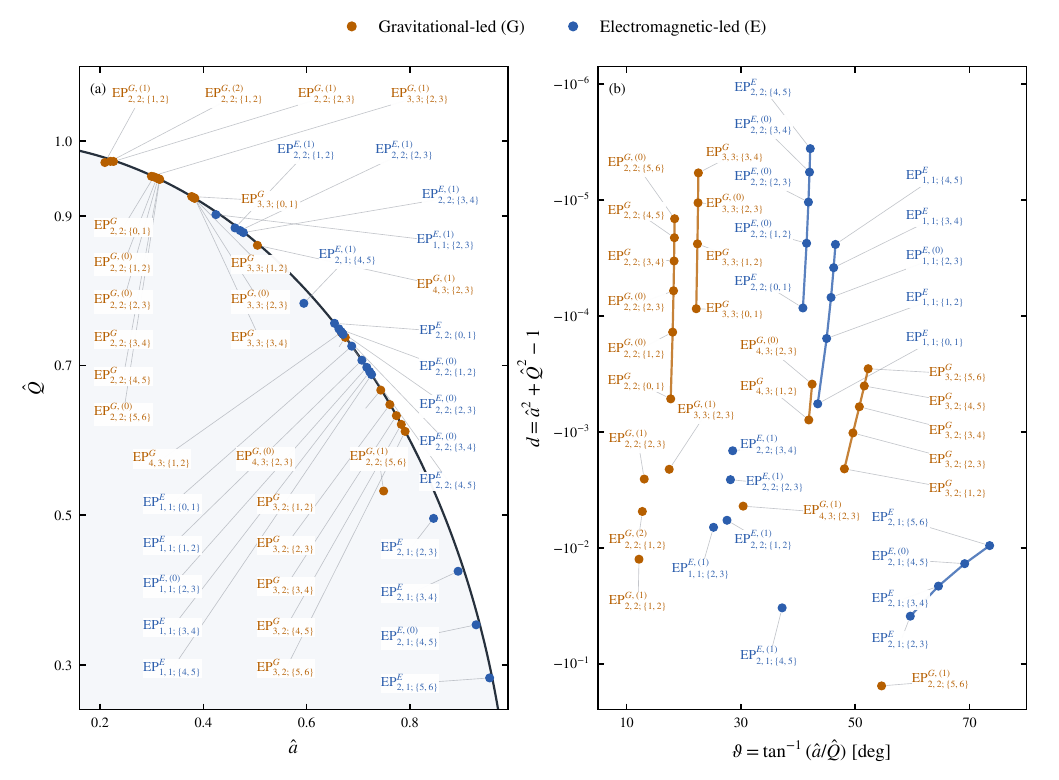}
\caption{Distribution of the EPs listed in Table~\ref{tab:EPs}. Panel (a) shows their locations in the $(\hat a,\hat Q)$ plane, where the solid curve marks the extremal boundary $\hat a^2+\hat Q^2=1$. Panel (b) shows the same EPs in the $(\vartheta,d)$ plane, where $\vartheta=\tan^{-1}(\hat a/\hat Q)$ and $d=\hat a^2+\hat Q^2-1$. Orange and blue points denote the gravitational-led and electromagnetic-led sectors, respectively.}\label{fig:EPdistribution}
\end{figure*}
%%%%%%%%%%%%%%%%%%%%%

Figure~\ref{fig:EPdistribution} shows the distribution of the 42 EPs listed in Table~\ref{tab:EPs}. Panel (a) gives their locations in the $(\hat a,\hat Q)$ plane, where the solid black curve marks the extremal boundary $\hat a^2+\hat Q^2=1$. Panel (b) displays the same EPs in the coordinates $(\vartheta,d_{\rm EP})$, with $\vartheta=\tan^{-1}(\hat a/\hat Q)$ and $d_{\rm EP}=\hat a_{\rm EP}^2+\hat Q_{\rm EP}^2-1$. Two features of this distribution are particularly noteworthy. First, both the gravitational-led and electromagnetic-led EPs exhibit ordered overtone-ladder structures, in which successive neighboring branches $\{n,n+1\}$ coalesce. These ladders appear in several angular sectors, indicating a systematic organization of the EP spectrum across the two perturbation branches. For example, in the electromagnetic-led $\ell=m=1$ sector, the overtone ladder takes the form
\begin{equation}
\begin{split}
\EPlabel{E}{1}{1}{0}{1}
&\rightarrow \EPlabel{E}{1}{1}{1}{2} \rightarrow \EPlabelj{E}{1}{1}{2}{3}{0} \\
&\rightarrow \EPlabel{E}{1}{1}{3}{4} \rightarrow \EPlabel{E}{1}{1}{4}{5} \rightarrow\cdots\,,
\end{split}
\label{eq:EP-ladder-l1}
\end{equation}
where the arrows indicate the ordering of the EPs with increasing overtone number. Along these ladders, increasing the overtone number is accompanied by increasing $\hat a_{\rm EP}$, decreasing $\hat Q_{\rm EP}$, and $d_{\rm EP}$ approaching zero. The higher-overtone members therefore lie progressively closer to the extremal KN boundary.

A second notable feature is that the base labels $(X,\ell,m,n_1,n_2)$ do not always specify a unique EP. For a fixed sector, angular mode, and overtone pair, the two-dimensional KN parameter space may contain several distinct coalescences. When this occurs, we assign $j=0$ to the EP belonging to the ordered overtone ladder and use $j=1,2,\ldots$ to distinguish the additional off-ladder EPs, ordered by increasing $\hat a_{\rm EP}$. For the gravitational-led $(2,2;\{1,2\})$ pair, three distinct EPs are found: $\mathrm{EP}^{G,(0)}_{2,2;\{1,2\}}$ belongs to the near-extremal ladder, whereas $\mathrm{EP}^{G,(1)}_{2,2;\{1,2\}}$ and $\mathrm{EP}^{G,(2)}_{2,2;\{1,2\}}$ lie away from this sequence. Similarly, the electromagnetic-led $(2,2;\{1,2\})$ pair possesses two EPs: $\mathrm{EP}^{E,(0)}_{2,2;\{1,2\}}$ lies on the near-extremal ladder, while $\mathrm{EP}^{E,(1)}_{2,2;\{1,2\}}$ is located off it. Thus, the KN parameter space contains not only ordered overtone ladders but also additional off-ladder coalescences carrying the same inherited overtone labels. The EP spectrum therefore has a more intricate organization than can be described by a single overtone ladder.

A closely related cascade of EPs has previously been found for massive scalar perturbations of near-extremal Kerr black holes, where successive overtone pairs become degenerate and the higher-overtone EPs accumulate toward the extremal limit~\cite{Cavalcante:2025abr}. Related exceptional-line and EP sequences have also been identified in the scalar Kerr--Newman spectrum~\cite{Cavalcante:2026vgr}. The present results differ in several important respects. First, the EP ladders arise in the fully coupled gravitoelectromagnetic perturbations of the KN background, with the two control parameters supplied intrinsically by the black hole spin and charge rather than by an additional scalar-field mass. Second, whereas the displayed scalar-field ladders in Kerr black hole begin with the fundamental--first-overtone pair $\{0,1\}$, the lowest identified member of a gravitoelectromagnetic ladder in KN black hole depends on the angular sector. For example, the gravitational-led $(\ell,m)=(3,2)$ ladder begins with $\{1,2\}$, while the electromagnetic-led $(2,1)$ ladder begins with $\{2,3\}$. Moreover, the occurrence of additional off-ladder EPs with the same mode labels shows that the coupled spectrum is not organized by a single cascade alone. To our knowledge, these results provide the first evidence for overtone ladders, together with repeated on- and off-ladder coalescences, in the fully coupled Einstein--Maxwell QNM spectrum.

\begin{figure*}[!t]
\centering
\includegraphics[width=\textwidth]{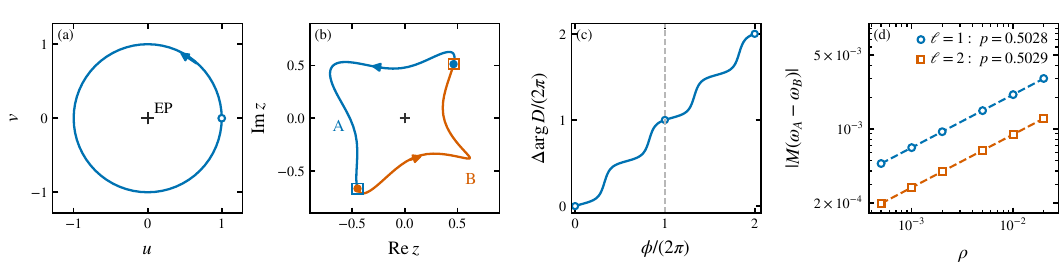}
\caption{Local diagnostics near representative EPs. Panels (a)--(c) refer to $\EPlabel{E}{1}{1}{0}{1}$: panel (a) shows the closed loop in the normalized $(u,v)$ parameter plane, panel (b) the corresponding trajectories of $z=10^3M(\omega-\omega_{\rm EP})$, and panel (c) the accumulated phase change of $D=[M(\omega_A-\omega_B)]^2$ over two circuits. Panel (d) shows
$|M(\omega_A-\omega_B)|$ versus $\rho=|\ln(\epsilon/\epsilon_{\rm EP})|$ for the electromagnetic-led $\ell=m=1$ and $\ell=m=2$ $\{0,1\}$ pairs; the dashed lines denote power-law fits.}\label{fig:EPdiagnostics}
\end{figure*}

Taking $\EPlabel{E}{1}{1}{0}{1}$ as a representative example, Fig.~\ref{fig:EPdiagnostics} examines the spectral structure around the coalescence and tests whether it exhibits the characteristic local branch-point structure of an EP. Using $\epsilon=\sqrt{1-\hat a^2-\hat Q^2}$ and $\vartheta=\tan^{-1}(\hat a/\hat Q)$, we parameterize a closed loop around the EP by $u=\ln(\epsilon/\epsilon_{\rm EP})/r_\epsilon=\cos\phi$ and $v=(\vartheta-\vartheta_{\rm EP})/r_\vartheta=\sin\phi$, with $0\leq\phi\leq4\pi$. After one circuit, the two tracked QNM frequencies exchange branches; only after a second circuit does each frequency return to its initial branch. The squared frequency difference $D=[M(\omega_A-\omega_B)]^2$ is invariant under the branch exchange, and its unwrapped phase increases by $2\pi$ per circuit. As an independent check of the local scaling of the frequency difference, we consider radial cuts in parameter space through the two EPs at their respective fixed values of $\vartheta=\vartheta_{\rm EP}$. Defining $\rho=|\ln(\epsilon/\epsilon_{\rm EP})|$, we fit
\begin{equation}
|M(\omega_A-\omega_B)|\propto\rho^p \,.
\end{equation}
The fits yield $p=0.5028$ and $0.5029$ for the electromagnetic-led $\ell=m=1$ and $\ell=m=2$ $\{0,1\}$ pairs, respectively. Both values are very close to $1/2$, consistent with the square-root scaling of the frequency difference near an EP. The branch exchange, phase winding, and square-root scaling therefore provide mutually consistent evidence for the local EP structure.

\begin{figure}[!t]
\centering
\includegraphics[width=\columnwidth]{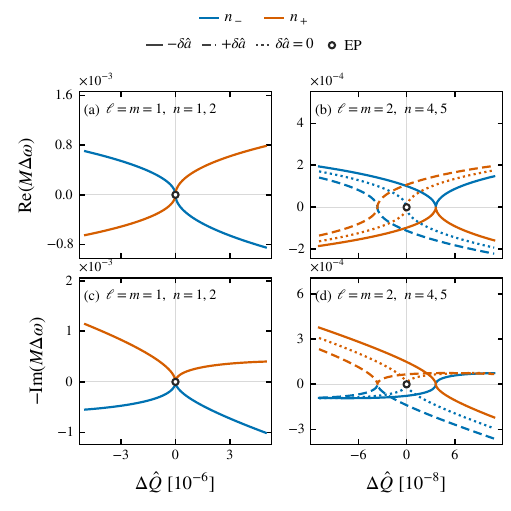}
\caption{Charge-dependent QNM spectra near $\EPlabel{E}{1}{1}{1}{2}$ (left column) and $\EPlabel{E}{2}{2}{4}{5}$ (right column). The upper and lower rows show $\operatorname{Re}(M\Delta\omega)$ and $-\operatorname{Im}(M\Delta\omega)$, respectively. Here $\Delta\omega=\omega-\omega_{\rm EP}$ and $\Delta\hat Q=\hat Q-\hat Q_{\rm EP}$, where $\omega_{\rm EP}$ and $\hat Q_{\rm EP}$ are the frequency and charge of the EP shown in the corresponding column. The left column is evaluated at $\hat a=\hat a_{\rm EP}$. In the right column, solid, dashed, and dotted curves correspond to $\hat a_{\rm EP}-\delta\hat a$, $\hat a_{\rm EP}+\delta\hat a$, and $\hat a_{\rm EP}$, respectively. Colors distinguish the two branches, with $n_-=n_1$ and $n_+=n_2$, and open circles mark the EPs.}\label{fig:localSpectrum}
\end{figure}

Figure~\ref{fig:localSpectrum} compares the charge-dependent QNM spectra in the vicinity of the lower-overtone $\EPlabel{E}{1}{1}{1}{2}$ and the higher-overtone $\EPlabel{E}{2}{2}{4}{5}$. For the lower-overtone pair, varying $\hat Q$ at fixed $\hat a=\hat a_{\rm EP}$ traces the two branches through the EP at $\Delta\hat Q=0$, where they coalesce. For the higher-overtone pair, changing the fixed spin by only $\delta\hat a=4.0192\times10^{-8}$ noticeably changes how the two QNM frequencies vary with charge. The two complex frequencies become equal only when $\hat a=\hat a_{\rm EP}$. When the spin is shifted slightly away from this value, their real or imaginary parts may cross separately, but the full complex frequencies remain different.

\section{Conclusions}\label{sec:conclusion}
%%%%%%%%%%%%%%%%%%%%%%%%
We have investigated the EFs and EPs of KN QNMs using the exact separated gravitoelectromagnetic perturbation equations. The EFs shown in Figs.~\ref{fig:Gmap} and~\ref{fig:EMmap} exhibit distinct dependencies on black hole spin and charge across perturbation sectors and overtones. For the four selected gravitational-led modes, the EF magnitudes decrease overall as the black hole spin increases at fixed charge, but the suppression rates differ substantially between overtones and reverse the relative ordering of the fundamental and first-overtone factors in the $\ell=m=2$ sector. In the electromagnetic-led sector, the EF magnitudes of the $(1,1,0)$, $(2,2,0)$, and $(2,2,1)$ modes also decrease with increasing spin, whereas the $(1,1,1)$ mode exhibits nonmonotonic behavior. Their charge dependence is likewise mode dependent. The relative magnitudes, phases, and complex-plane trajectories further show that rotation can reorganize the excitation properties of different overtones. These results demonstrate that the excitation properties of KN QNMs depend on both the black hole parameters and the mode labels, rather than following a universal trend.

We have also identified 42 EPs in the coupled gravitoelectromagnetic QNM spectrum, with 23 in the gravitational-led sector and 19 in the electromagnetic-led sector, as summarized in Table~\ref{tab:EPs} and Fig.~\ref{fig:EPdistribution}. Their distribution reveals two notable forms of spectral organization. First, both sectors contain ordered overtone ladders formed by successive coalescences of neighboring branches $\{n,n+1\}$, with the higher-overtone members moving progressively closer to the extremal KN boundary. The lowest identified member of a ladder depends on the angular sector and need not involve the fundamental mode. Second, the same sector, angular mode, and overtone pair may correspond to several distinct EPs. Thus, the EP spectrum contains both systematic overtone sequences and repeated coalescences that cannot be described by a single ladder alone. These results extend the overtone-ladder structure previously observed for probe scalar fields~\cite{Cavalcante:2025abr} to the fully coupled Kerr--Newman QNM spectrum. Figs.~\ref{fig:EPdiagnostics} and \ref{fig:localSpectrum} confirm the characteristic square-root branch structure of the selected EPs and reveal the strong parameter sensitivity of higher-overtone QNMs. Together, these results demonstrate the rich non-Hermitian structure of the KN QNM spectrum.

Several directions deserve further investigation. A broader search over additional angular sectors and higher overtones is needed to determine how far the overtone ladders extend and to clarify how the accompanying off-ladder EPs are distributed relative to these ladders. An analytic treatment of the near-extremal regime may clarify why the higher-overtone members approach the extremal boundary and whether these sequences are related to photon-sphere, damped, or near-horizon mode families~\cite{Dias:2021yju,Dias:2022oqm}. An important extension is to study QNM excitation by specified external sources, such as a compact object plunging into a KN black hole from the innermost stable circular orbit. Combining the source-independent EFs with source-dependent contributions and the reconstruction of physical gravitational and electromagnetic perturbations would provide a more complete description of black hole ringdown signals. It would also be valuable to formulate the resonant response directly near an EP, where the coalescence of QNM frequencies and eigenfunctions may require a generalized modal description beyond a sum of isolated simple modes~\cite{PanossoMacedo:2025xnf}. Such investigations would deepen our understanding of how gravitational--electromagnetic coupling shapes the dynamical response and spectral properties of rotating, charged black holes.

\section{Acknowledgments}
The work is funded by the National Natural Science Foundation of China with Grants No.~12405059 and No.~12505087, the Startup Fund for Advanced Talents of Nanchang University with Grant No. 28170256, Key Program of the Natural Science Foundation of Jiangxi Province under Grant No. 20232ACB201008 and the Ganpo Talent Support Program with No. gpjc20260105.

\appendix
\section{Angular formulation}\label{app:angular-basis}
We compute KN QNMs using the spectral method of Ref.~\cite{Destounis:2026ljd}, based on the exact separated gravitoelectromagnetic equations derived in Ref.~\cite{Hintz:2026jiw}. The method combines an angular Galerkin expansion with radial Chebyshev collocation to solve simultaneously for the complex frequency $\omega$ and the separation constant $\lambda$.
We outline the formulation and numerical procedure below.

We use the mode convention and the angular functions $W$ and $T$ defined in Sec.~\ref{subsec:KN_separated_perturbations} and Eq.~\eqref{eq:KN_WT}. The label $\varsigma=\pm1$ distinguishes the two spin descriptions. In the two-component basis used in Ref.~\cite{Destounis:2026ljd}, the restricted angular operator is
\begin{equation}
\mathcal A=
\begin{pmatrix}
i\partial_\theta&T-iW\\
T+iW&-i\partial_\theta
\end{pmatrix}.
\end{equation}
The constant transformation
\begin{equation}
u=H\chi,\qquad
H=\frac{1}{\sqrt2}
\begin{pmatrix}1&1\\1&-1\end{pmatrix}
=H^{-1}
\end{equation}
brings its eigenvalue equation into the form of~\eqref{eq:KN_angular_matrix}. We denote the resulting matrix differential operator by $\mathcal B(\omega)$. This change of basis preserves both the eigenvalues and the regularity conditions at the poles. It remains valid for complex $\omega$, although the angular operator is then generally non-self-adjoint.

At $\omega=0$, the rotation-dependent terms in the angular equations vanish, and the eigenvalues for $\ell\geq2$ are given by Eq.~\eqref{eq:KN_lambda_seed}. The label $b$ identifies an eigenvalue branch continued from this limit and is distinct from the spin label $\varsigma$. The correspondence between $b$ and the gravitational-led and electromagnetic-led sectors is established through the Kerr limit discussed below.

\section{Sector identification in the Kerr limit}\label{app:Q-to-0}

The apparent divergence of $\lambda$ as $Q\to0$ can be separated from its finite part by writing
\begin{equation}
\lambda=b\left(\frac{3M}{2Q}+\frac{Q}{M}\widetilde\lambda\right).
\label{eq:regular-lambda}
\end{equation}
Following Ref.~\cite{Destounis:2026ljd}, introduce $q=Q/M$ and rescale the angular and separated radial amplitudes according to
\begin{equation}
\begin{gathered}
\chi=D_b\widetilde\chi,\qquad z=D_b\widetilde z,\\
D_+=\operatorname{diag}(q,1),\qquad
D_-=\operatorname{diag}(1,q).
\end{gathered}
\label{eq:sector-rescaling}
\end{equation}
The branches of interest have finite $\widetilde\lambda$ as $q\to0$. These rescalings yield angular and radial equations with finite coefficients in this limit. At nonzero charge, the two rescaled angular problems are reparameterizations of the same spectrum; their distinct Kerr limits identify the two sectors.

The Kerr limit is taken after the sector-dependent field rescalings and the corresponding redefinition of the radial master variable described in~\cite{Destounis:2026ljd}. It cannot be obtained by directly setting $Q=0$ in Eq.~(13), whose coefficients are singular in this limit. In particular, the limiting radial equations reduce to the Teukolsky equations with spin $s=2\varsigma$ for $b=+1$ and $s=\varsigma$ for $b=-1$. With the convention $\Lambda_s={}_sA_{\ell m}+a^2\omega^2-2am\omega$, their separation constants satisfy
\begin{equation}
\Lambda_s=
\begin{cases}
3\widetilde\lambda-2-2\varsigma,&b=+1,\\
3\widetilde\lambda+1-\varsigma,&b=-1.
\end{cases}
\label{eq:kerr-separation-constant}
\end{equation}
Thus $b=+1$ and $b=-1$ label the gravitational-led and electromagnetic-led modes, respectively. Away from $Q=0$, both sectors retain gravitational--electromagnetic coupling.

\section{Spectral discretization}\label{app:radial-angular-discretization}
\subsection{Angular projection}\label{app:angular-galerkin}
For each angular component $k=1,2$, set $z=\cos\theta$ and $j_{\min,k}=\max(|m|,k)$, and use the pole exponents $\alpha_k$ and $\beta_k$ from~\eqref{eq:KN_pole_exponents}, with the component index $s$ replaced by $k$. We approximate the regular solution by
\begin{equation}
\chi_k(\theta)=
\sum_{j=j_{\min,k}}^L c_{jk}\Phi_{jk}(\cos\theta),
\label{eq:angular-expansion}
\end{equation}
where $L$ is the angular cutoff and
\begin{equation}
\begin{split}
\Phi_{jk}(z)={}&\mathcal N_{jk}
(1-z)^{\alpha_k/2+1/4}
(1+z)^{\beta_k/2+1/4}\\
&\times P_{j-j_{\min,k}}^{(\alpha_k,\beta_k)}(z).
\end{split}
\label{eq:angular-jacobi-basis}
\end{equation}
The Jacobi polynomials provide the regular part of each component, while the factors impose the required behavior at both poles. The power expansions with half-integer arise from the rescaled angular variables.

The basis is orthonormal with respect to
\begin{equation}
\langle f,g\rangle
=\int_0^\pi f^*(\theta)g(\theta)\,d\theta
=\int_{-1}^1\frac{f^*(z)g(z)}{\sqrt{1-z^2}}\,dz.
\label{eq:angular-inner-product}
\end{equation}
Writing $d=j-j_{\min,k}$, the normalization is
\begin{equation}
\mathcal N_{jk}^2=
\frac{(2d+\alpha_k+\beta_k+1)d!\,
\Gamma(d+\alpha_k+\beta_k+1)}
{2^{\alpha_k+\beta_k+1}
\Gamma(d+\alpha_k+1)\Gamma(d+\beta_k+1)}.
\end{equation}
Here $k$ denotes the component and $d$ the polynomial degree; neither is the radial overtone index $n$.

Let $\Psi_{jk}=\Phi_{jk}e_k$, with $e_k$ the two-component coordinate vectors. Since $\mathcal B(\omega)=\mathcal B_0+\omega\mathcal B_1$, Galerkin projection gives
\begin{equation}
G(\omega,\lambda)c=0,\qquad
G=G_0+\omega G_1-\lambda I,
\label{eq:angular-galerkin-matrix}
\end{equation}
where
\begin{equation}
(G_p)_{jk,j'k'}=
\langle\Psi_{jk},\mathcal B_p\Psi_{j'k'}\rangle,
\qquad p=0,1.
\end{equation}
The vector inner product includes a sum over components. Matrix elements are evaluated by Gauss--Legendre quadrature with $4L+16$ nodes in $z$, retaining the measure in~\eqref{eq:angular-inner-product}; basis derivatives are evaluated analytically. The resulting matrix has dimension
\begin{equation}
N_{\rm ang}=\sum_{k=1}^2(L-j_{\min,k}+1).
\end{equation}
All basis functions allowed by $j_{\min,k}$ are retained, including the additional lower-order function when $|m|=1$. Although $G$ is affine in $\omega$, the selected eigenvalue $\lambda(\omega)$ generally depends nonlinearly on frequency.

\subsection{Radial compactification}

We start from the radial equation~\eqref{eq:KN_radial_ABC}, with the coefficients in~\eqref{eq:KN_radial_coefficients} and the background and radial functions defined in Sec.~\ref{subsec:KN_separated_perturbations}. The horizon and infinity conditions are those of Sec.~\ref{subsec:sol}. For $Q\omega\neq0$, we use the coordinate $x$ introduced in Sec.~\ref{subsec:sol} and extract the outgoing asymptotic factor with horizon normalization:
\begin{equation}
\begin{aligned}\label{eq:appendix-outgoing-factor}
w(r)&=F(r)f(x) \,, \\
F(r)&=e^{i(2\omega+\xi_0)(r-r_+)}\left( \frac{r}{r_+}\right)^{4i M\omega} \,.
\end{aligned}
\end{equation}
The horizon and infinity correspond to $x=0$ and $x=1$, respectively, and $F(r_+)=1$.

To transform the derivatives, define
\begin{equation}
\tau=\frac{dx}{dr}=\frac{r_+}{r^2},
\qquad
g=\frac{F'}{F}=i(2\omega+\xi_0)+\frac{4iM\omega}{r}.
\end{equation}
Then
\begin{equation}
\frac{w''}{F}
=\tau^2 f_{xx}+(\tau'+2g\tau)f_x+(g'+g^2)f,
\end{equation}
where $\tau'=-2r_+/r^3$ and $g'=-4iM\omega/r^2$. After multiplying the transformed equation by $1-x$, we obtain
\begin{equation}
c_2f_{xx}+c_1f_x+c_0f=0,
\label{eq:appendix-compact-equation}
\end{equation}
with
\begin{align}
c_2&=(1-x)A\tau^2,\\
c_1&=(1-x)\bigl[A(\tau'+2g\tau)+B\tau\bigr],\\
c_0&=(1-x)\bigl[A(g'+g^2)+Bg+C\bigr].
\label{eq:appendix-compact-coefficients}
\end{align}
All radial functions are evaluated at $r=r_+/(1-x)$. The large-$r$ cancellations are performed analytically before evaluating the coefficient limits at $x=1$.

\subsection{Radial collocation and boundary conditions}

We discretize Eq.~\eqref{eq:appendix-compact-equation} on $N+1$ Chebyshev--Lobatto points,
\begin{equation}
x_j=\frac{1-\cos(j\pi/N)}{2},
\qquad j=0,\ldots,N.
\end{equation}
If $D$ denotes the differentiation matrix with respect to $x$, the radial matrix is
\begin{equation}
R(\omega,\lambda)
=\operatorname{diag}(c_2)D^2
+\operatorname{diag}(c_1)D
+\operatorname{diag}(c_0).
\label{eq:appendix-radial-matrix}
\end{equation}
The endpoint rows use the finite limiting coefficients. Generically, $c_2(0)=c_2(1)=0$, and these rows reduce to first-order compatibility relations.

These relations implement the limiting differential equation, but smoothness alone does not establish the radiative boundary conditions for complex frequencies. In particular, after removing $F$, the competing asymptotic solution contains an $e^{-2i\omega r}$ factor and can be smooth and exponentially
small at infinity for damped modes. As emphasized in Ref.~\cite{Destounis:2026ljd}, identifying physical QNMs therefore requires the prescribed asymptotic branch, together with resolution studies and independent frequency comparisons.

\section{Simultaneous solution and mode tracking}
\label{app:joint-eigenvalues}

A QNM corresponds to a common pair $(\omega,\lambda)$ for which both $G(\omega,\lambda)$ and $R(\omega,\lambda)$ admit nonzero null vectors. We locate these pairs using the normalized residual construction of Ref.~\cite{Destounis:2026ljd}. For either matrix $\mathcal M=G$ or $R$, fixed row scaling is applied, and one equation is replaced by a normalization condition $v_k=1$. If row $j$ is omitted, this construction is
\begin{align}
\widehat{\mathcal M}
&=\mathcal M+e_j(e_k^T-e_j^T\mathcal M),\\
\widehat{\mathcal M}v&=e_j,\qquad
F_{\mathcal M}(\omega,\lambda)=e_j^T\mathcal Mv,
\label{eq:appendix-normalized-residual}
\end{align}
where $e_j$ and $e_k$ are coordinate vectors of the appropriate dimension. Provided $\widehat{\mathcal M}$ is invertible, $F_{\mathcal M}=0$ restores the omitted equation and yields a nontrivial null vector of $\mathcal M$. The normalization fixes an arbitrary eigenvector amplitude and adds no physical boundary condition.

The replacement row and normalization component are selected from the left and right singular vectors associated with the smallest singular value at the initial estimate. These choices, the row scalings, and the residual scalings remain fixed during each local search. We solve
\begin{equation}
F_G(\omega,\lambda)=0,\qquad
F_R(\omega,\lambda)=0
\label{eq:appendix-joint-conditions}
\end{equation}
as four real equations for the real and imaginary parts of $\omega$ and $\lambda$, using bounded nonlinear least squares followed by local root refinement.

For $\ell\geq2$, the angular branch is initialized at $\omega=0$ using~\eqref{eq:KN_lambda_seed}. The $\ell=1$ electromagnetic-led branch is identified within the regular $\omega=0$ angular problem. The selected eigenpair is continued to the trial frequency along $\omega(\eta)=\eta\omega_{\rm trial}$,
$0\leq\eta\leq1$, choosing successive eigenvectors by their largest normalized overlap. The angular branch assignment is checked again at the converged frequency.

During a scan in $(a,Q)$, solutions at neighboring parameter values provide initial estimates for the next calculation. The radial overtone label is inherited by continuation from an identified reference mode; it is not determined by the angular labels alone. Independent changes of $N$ and $L$ test the radial and angular truncations, respectively. Comparison with published QNM frequencies provides a further validation of the computed roots, as discussed in Appendix~\ref{app:numer:error}.

\section{Numerical check of our method}\label{app:numer:error}

\begin{table}[!thbp]
\centering
\small
\setlength{\tabcolsep}{3pt}
\renewcommand{\arraystretch}{1.12}
\resizebox{\columnwidth}{!}{%
\begin{tabular}{@{}ccc@{}}
\hline\hline
\multicolumn{3}{c}{
QNM benchmarks: Ref.~\cite{Destounis:2026ljd}}\\
\hline
$(\hat a,\hat Q;\,b)$ & $M\omega$ & $\epsilon_{\rm QNM}$\\
\hline
$(0.2,0.4;\,+1)$
& $0.4099175-0.088621\,\mathrm{i}$
& $1.854\times10^{-5}$\\
$(0.2,0.4;\,-1)$
& $0.5150483-0.0955546\,\mathrm{i}$
& $3.209\times10^{-5}$\\
\hline
\multicolumn{3}{c}{
EP benchmarks: Ref.~\cite{Arcos:2026eis}}\\
\hline
$(\hat a,\hat Q;\,b)$ & $M\omega_{\rm EP}$ & $\epsilon_{\rm EP}$\\
\hline
$(0.304116510,0.952362666;\,+1)$
& $0.564398227-0.028304554\,\mathrm{i}$
& $9.537\times10^{-5}$\\
$(0.461271757,0.883991733;\,-1)$
& $0.793226261-0.144859314\,\mathrm{i}$
& $1.985\times10^{-4}$\\
\hline\hline
\end{tabular}%
}
\caption{Comparison of our QNM and EP frequencies with published results. Both QNMs have $\ell=m=2$ and $n=0$; reference values are from Table~I of Ref.~\cite{Destounis:2026ljd}. The EP data are from Table~\ref{tab:EPs} and compared with $\mathrm{EP}_0$ and $\mathrm{EP}_e$ in Table~I of Ref.~\cite{Arcos:2026eis}. The relative differences are defined in Eq.~\eqref{eq:relative:diff}.}\label{tab:QNM:error}
\end{table}

Table~\ref{tab:QNM:error} compares our QNM and EP frequencies with published results. The QNMs are compared at the same background parameters and with the same mode labels. For the EPs, we use our results from Table~\ref{tab:EPs} and compare the frequencies at the EP locations found in each study. The relative frequency difference is \begin{equation}\label{eq:relative:diff}
\epsilon_i =
\frac{|M\omega_{r,i}-M\omega_{p,i}|}
{|M\omega_{p,i}|},
\qquad i\in\{\mathrm{QNM},\mathrm{EP}\},
\end{equation}
where $p$ and $r$ denote our result and the reference result, respectively. For the EPs, we use the published central values.

\bibliography{main}
\end{document}